# Diluted magnetism in Mn–doped SrZnO$_2$ single crystals

M R Rahman[1,2], B Koteswararao[1,3], S H Huang[1], Kee Hoon Kim[3] and F C Chou[1,4,a]

[1]*Center of Condensed Matter Sciences, National Taiwan University, Taipei 10617, Taiwan*

[2] *Department of Metallurgical and Materials Engineering, National Institute of Technology Karnataka, Surathkal, 575025, India*[3]*CeNSCMR, Department of Physics and Astronomy, Seoul National University, Seoul 151-747, Republic of Korea*

[4]*National Synchrotron Radiation Research Center, HsinChu 30076, Taiwan, Republic of China*

E–mail: fcchou@ntu.edu.tw

**Abstract.** We have investigated the magnetic properties of Mn– and Cu– substituted SrZnO$_2$ single crystals (SrZn$_{1-x}$Mn$_x$O$_2$ and SrZn$_{1-x}$Cu$_x$O$_2$). We observed signatures of weak ferromagnetism as a sharp increase of magnetic susceptibility below 5 K even in the low–percentage ($x= 0.01$) of Mn–substituted single crystals. Magnetic susceptibility data measured parallel or perpendicular to the *ab*-plane yield anisotropic behavior with Curie–Weiss temperature of about -320 K and -410 K, respectively, suggesting the presence of strong antiferromagnetic couplings among Mn atoms at high temperatures, similar to the Mn–doped ZnO and Fe–doped BaTiO$_3$ samples. In contrast, the SrZn$_{0.99}$Cu$_{0.01}$O$_2$ crystal shows paramagnetic behavior down to 2 K.

a)Electronic mail: fcchou@ntu.edu.tw

## I. Introduction

Significant efforts have been devoted over the past decade in exploring the materials having both semiconducting and ferromagnetic properties at room temperature.[1,2] Those materials with simultaneous ferromagnetic and semiconducting property are vital for realizing spintronic devices in which the electronic spin is used in addition to its charge for both memory and processing.[3] Research activities toward this direction have been focused on the study of dilute magnetic semiconductors (DMS) considerably.[4] Initial researches on the DMS materials were dealing with the systems such as (Cd,Mn)Te,[5] (Ga,Mn)As,[6] etc. with ferromagnetic (FM) ordering only below 100 K. Later the observation of room temperature (RT) ferromagnetism in transition-metal doped ZnO[7] attracted more interests in this area. More FM materials at RT such as doped $TiO_2$, $SnO_2$, $In_2O_3$, etc. were also found.[8,9]

Theoretical predictions based on the Zener model[10] suggest that ZnO has a high Curie temperature (above RT) for 5% of Mn–substitution at the Zn–site and hole concentration can be raised up to $3.5 \times 10^{20}$ per $cm^3$, which facilitates a transparent magnet for application in optoelectronic devices. However, the ferromagnetism in ZnO is extremely sensitive to synthesis methods and conditions.[11–13] In addition, ZnO based DMS is prone to the spurious magnetic signals such as O–vacancies, defects and grain boundaries.[14,15] It is also reported that ferromagnetism in this material could be non–intrinsic.[16,17] The origin of room temperature FM ordering in this material is still controversial and thus under both theoretical and experimental investigations.

It would be indeed necessary to study the DMS materials of interest in a single crystal form to overcome the issues related to the sample quality. Moreover, in order to clarify the intrinsic mechanism of the ferromagnetism in the DMS materials, a parallel study of the dilute magnetism in a ZnO-based wide band gap semiconductors is highly desirable. Recently it is reported that $SrZnO_2$ has a wide band gap of 3.4 eV which is close to that of ZnO.[18] The photoluminescence (PL) study on $SrZnO_2$ doped with different transition metals and rare–earth metals[19,20] propose that it can be a strong candidate for a spin–LED. However, there are no magnetic property studies reported for the transition-metal- or rare–earth-doped $SrZnO_2$ system.

Herein, we report the magnetic properties of $SrZn_{0.99}Mn_{0.01}O_2$ and $SrZn_{0.99}Cu_{0.01}O_2$ single crystals grown by the optical floating zone method. Ferromagnetism is observed at $T_C = 5$ K in the $SrZn_{0.99}Mn_{0.01}O_2$ single crystal, despite the presence of relatively large antiferromagnetic (AF) coupling ($\theta_{cw} \sim -320$ K) between $Mn^{2+}$ ions at high temperature. On the other hand, paramagnetic behavior is observed in the Cu–doped $SrZnO_2$ single crystals within the temperature range of 2–300 K.

## II. Experimental

Polycrystalline samples of $SrZn_{1-x}M_xO_2$ ($x = 0$ and 0.01, M = Mn and Cu) were synthesized by the conventional solid–state reaction method by mixing the stoichiometric amount of precursors $SrCO_3$, ZnO, $MnCO_3$ and CuO (99.99 % purity). The mixture was well ground in an acetone medium using an agate mortar and a pestle and then pelletized. The pellets were fired from 800–1100 °C for 100 hours with several intermediate grindings to improve homogeneity at such a small doping level ($x = 0.01$). These pellets were then quenched from 1100 °C into liquid $N_2$ to avoid formation of impurity phases. An optical floating zone furnace (Crystal System Inc.) was used to grow the single crystals of $SrZn_{1-x}M_xO_2$. Polycrystalline rods of 3.5 mm diameter were sintered at 1100 °C and used as feed and seed rods for the optical floating zone growth. A constant oxygen gas pressure of 0.140 MPa was maintained in the chamber during the growth. The feed and seed rods were counter–rotated at a speed of 30 rpm and the crystals were pulled at a rate of 5 mm/hr. X–ray diffraction (XRD) was used to check the single phase of powders and single crystals using the high resolution diffractormeter (Bruker, Advance D8). The magnetization measurements were carried out using the SQUID–VSM (Quantum Design) from 2 to 300 K and in the magnetic fields up to 7 T.

### III. Results and discussion

#### A. Crystal structure and XRD

The crystal structure of $SrZnO_2$ was first determined by Schnering and Hoppe in 1961.[21] The $SrZnO_2$ has an orthorhombic structure in the space group *Pnma* (No. 62). The structure can also be viewed as a quasi 2-dimensional (2D) structure built with edge–shared $ZnO_4$ tetrahedrons in the *ab*–plane, and the voids between layers are filled with Sr, as shown in Fig. 1. The powder XRD patterns for the $SrZn_{1-x}M_xO_2$ ($x = 0$, 0.01 and M = Mn, Cu) samples are matched with the standard JCPDS (file # 41–0551) pattern of $SrZnO_2$ with the lattice constants of $a = 3.383$ (5) Å, $b = 5.895$ (5) Å and $c = 11.361$ (5) Å, respectively. No trace of other secondary phases is observed in the samples up to the 1% of dopant concentration, whereas we observed a secondary phase when the Mn doping level increased to 2 %, which is consistent with the previous report [19]. It is confirmed that the solubility limit for Mn–substitution in the $SrZnO_2$ is less than 2%. The single crystals have an easy cleave plane in the (00*l*) direction as confirmed by XRD (Fig. 2). The lattice parameters of Mn– and Cu–doped samples showed no significant change in their values compared to the pure system, which could be due to the low–doping concentration. However, the color of the $SrZnO_2$ has easily changed by a small percent of Cu or Mn–substitution. The color of the crystal is uniform, which indicates that the Mn or Cu atoms are rather distributed homogeneously throughout the crystal, which could preclude the possibility of ferromagnetism originated from the Mn–clusters.

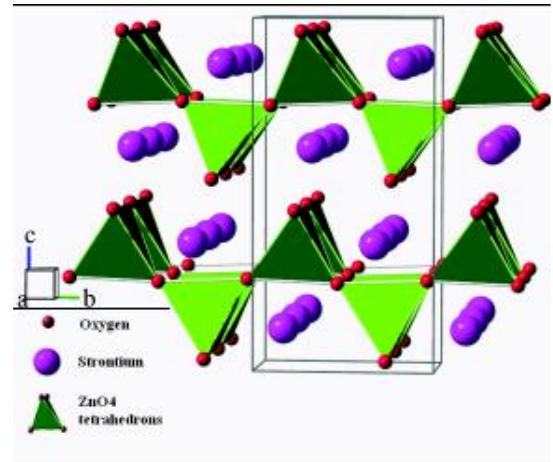

FIG. 1.(color online) Crystal structure of $SrZnO_2$ viewed along the crystallographic *a*–axis. The $ZnO_4$ unit, Sr, and O atoms are represented by green, violet and red colors respectively.

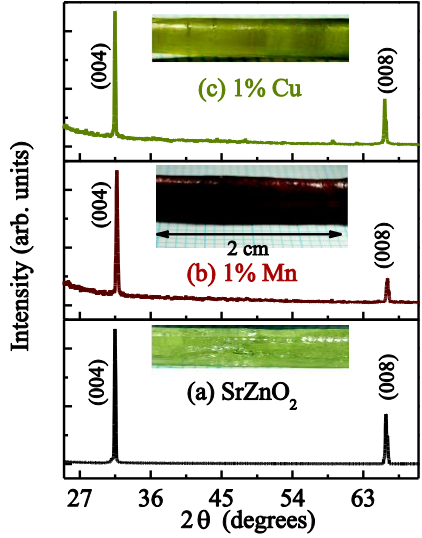

FIG. 2. (color online) XRD patterns of the cleaved (00*l*) plane of (a) SrZnO$_2$ (b) 1% Mn–doped and (c) 1% Cu–doped crystals. The color of the crystal changes drastically even after only 1% of doping.

## B. Magnetic analysis and discussion
### 1. Pure sample

Figure 3 shows the magnetic susceptibility ($\chi_{pure} \equiv M/H$) versus temperature plot for the SrZnO$_2$ sample from 2 to 300 K measured in a magnetic field ($H$) of 5 kOe. The measurement is performed on the single crystals along both directions of $H//ab$ (in–plane) and $H\perp ab$ (out–of–plane). The in–plane and out–of–plane anisotropy is absent for the pure sample. The data of $\chi_{pure}$ is negative down to 2 K, which suggests that the crystal remains diamagnetic. A small upturn observed below 30 K may be due to a very small fraction of paramagnetic impurities and/or the oxygen vacancies. The data is fitted to the Curie–Weiss form with a constant ($C/(T - \theta_{cw}) + \chi_0$) in the $T$–range from 2 to 300 K, as listed in Table I. The obtained $\chi_0$ value is about $-(5\pm0.05)\times10^{-5}$

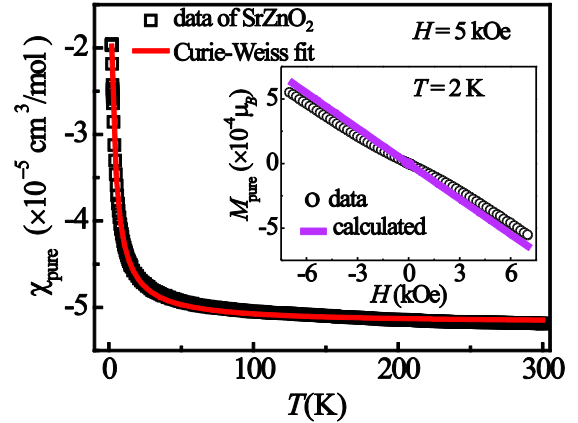

FIG.3.(color online)The spin susceptibilities χ of SrZnO$_2$ are plotted versus temperature and fitted to the Curie–Weiss law (red line). The inset shows the $M$ vs $H$ curve for the SrZnO$_2$ at 2 K in the $H$–range from –7 T to +7 T.

cm$^3$/mol, which is nearly equal to the value of calculated diamagnetic susceptibility ($\chi_{dia}$) $-5.1\times10^{-5}$ cm$^3$/mol from the individual ions.[22] The yielded Curie constant (C) is about $1.1\times10^{-4}$ cm$^3$ K/mol Cu$^{2+}$ ion, which corresponds to the 0.35 % of $S = ½$ moments. The plot of magnetization vs magnetic field ($M_{pure}$ vs $H$) for SrZnO$_2$ at 2 K is shown in the inset of Fig. 3, which deviates slightly from the calculated magnetization using $\chi_0$, and supports the presence of small percentage of paramagnetic impurities in the pure sample.

### 2. SrZn$_{0.99}$Mn$_{0.01}$O$_2$ sample

The measured χ of SrZn$_{0.99}$Mn$_{0.01}$O$_2$ (SZM) crystal ($\chi_{SZM}$) is expected to have contributions from the diamagnetic susceptibility, paramagnetic impurities, and Mn$^{2+}$ spins. The contribution coming purely from Mn$^{2+}$ ions can be isolated by subtracting the data for the pure SrZnO$_2$ from the $\chi_{SZM}$ data. Inverse of the obtained data ($\chi_{SZM} - \chi_{pure}$)$^{-1}$ is plotted in Fig. 4. There is a significant difference between the in–plane and out–of–plane susceptibility below 250 K. The observed relative difference ($\Delta\chi/\chi$) of 40 % between the $H//ab$ and $H\perp ab$ data is larger than the common difference for the Van Vleck susceptibilities (usually less than about

~10 %), which suggests that the Mn–doped system has a large magnetic anisotropy. The data of $H//ab$ and $H\perp ab$ are fitted using inverse Curie–Weiss law $((T-\theta_{cw})/C)$ in the $T$–range from 100 to 300 K. The yielded fitting parameters are listed in the Table I. The yielded values of Curie constant ($C$) give the effective moments ($\mu_{eff}$) of ~5.8 $\mu_B$/Mn (for $H//ab$) and 5.5 $\mu_B$/Mn (per Mn) (for $H\perp ab$), which are in nearly agreement with the expected value of 5.92 $\mu_B$ for $3d^5$ ($S = 5/2$ and $g = 2$) spins. The obtained values of $\theta_{cw}$ for the $H//ab$ plane and $H\perp ab$ plane are about $-(410\pm10)$ K and $-(320\pm5)$ K, respectively. The difference between $\theta_{cw}$ for $H//ab$ and $H\perp ab$ might be associated with the presence of spin anisotropy, presumably related to either quasi-2D crystallographic structure or local anisotropic AF interactions. In any case, this large negative $\theta_{cw}$ value suggests the presence of strong AF coupling between the Mn atoms even for such a diluted doping level (1 %). The average antiferromagnetic exchange coupling ($J_{avg}$) between Mn atoms is related to the $\theta_{cw}$ by $\frac{J_{avg}}{k_B} = \frac{3}{2}\frac{\theta_{CW}}{zxS(S+1)}$,[23] where $zx$ is the average number of the nearest-neighbor Mn atoms surrounded by a Mn atom and $z = 4$ for SrZnO$_2$ system. The obtained $J_{avg}/k_B$ value is about $-15$ K.

TABLE I. The fitting parameters of Curie–Weiss law fitting for $\chi_{pure}$ and the spin susceptibilities purely coming from Mn/Cu contribution.

| Sample | Direction | $C$(cm$^3$/molMn or Cu) | $\theta_{cw}$ (K) |
|---|---|---|---|
| SrZnO$_2$ | $H//ab$ and $H\perp ab$ | $1.1\times10^{-4}$ | $-(2\pm0.5)$ |
| 1%Mn | $H//ab$ | $4.21\pm0.05$ | $-(410\pm20)$ |
|  | $H\perp ab$ | $3.83\pm0.05$ | $-(320\pm10)$ |
| 1%Cu | $H//ab$ | $0.25\pm0.02$ | $-(3\pm0.5)$ |
|  | $H\perp ab$ | $0.37\pm0.03$ | $-(2\pm0.5)$ |

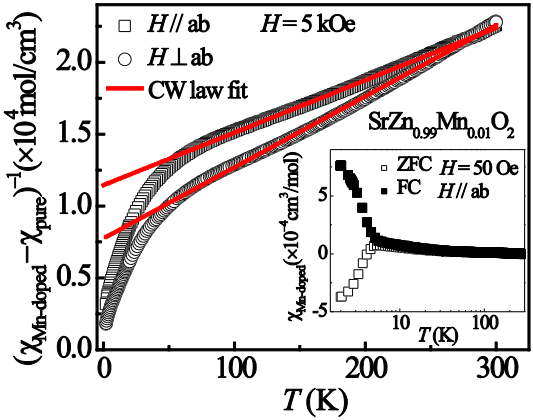

FIG. 4. (color online) The inverse of spin susceptibilities, $(\chi_{SZM}-\chi_{pure})^{-1}$, for single crystal SrZn$_{0.99}$Mn$_{0.01}$O$_2$ along $H//ab$ (open square) and $H\perp ab$ (open circle). Red lines represent the fit to the Curie–Weiss law. The inset shows the data of $\chi_{SZM}$ vs. $T$ in semi–log scale for the $H//ab$ plane of SZM single crystal at 50 Oe.

These diluted Mn moments order ferromagnetically at 5 K ($=T_C$), as evidenced from the ZFC and FC splitting in the in–plane data ($\chi_{SZM}$) (inset of Fig. 4). Similar features were also observed in the $H\perp ab$ data (*not shown here*). The ZFC and FC splitting is generally observed in spin–glass–like materials. Generally, the spin–glass behavior produces a weak ZFC and FC splitting, while our data shows a large bifurcation from a negative (for ZFC) to a positive value (for FC) with nearly same magnitude. This indicates that for ZFC, the sample is ordered ferromagnetically below 5 K with the magnetization direction opposite to the field direction.

In order to confirm the weak FM nature below $T_C = 5$ K, we performed the $M$-$H$ loop measurements at 2 K. Figure 5 shows the curve of ($M_{SZM}$–$M_{pure}$) vs $H$ at 2 K in the $H$-range from –250 Oe to 250 Oe for both $H//ab$ and $H\perp ab$ directions. The appreciable hysteresis loops are observed, which is an evidence of the weak ferromagnetism in the bulk SZM. At 2 K, it has the coercive field ($H_c$) of ~ 84 Oe and ~ 40 Oe and the remnant magnetization ($M_r$) $5.18\times10^{-4} \mu_B$/Mn and $4.34\times10^{-4} \mu_B$/Mn, for $H//ab$ and $H\perp ab$ plane, respectively. The inset of Fig. 5 shows $M_{SZM}$–$M_{pure}$ vs $H$ plot at 2 K up to 7 T for SZM. The linear $M(H)$ behavior above 250 Oe indicates that the FM signal below 5 K is a weak ferromagnetic transition. The large AF interactions at high-$T$ and weak ferromagnetic signature with spin anisotropy at low-$T$ suggests that this magnetic ordering might be associated with the canted antiferromagnetic ordering or weak ferromagnetic ordering with yet strong, anisotropic AF fluctuations.

There have been many studies to explore the origin of the FM signatures in the DMS materials. Most commonly accepted cause for the FM coupling is the carrier mediated exchange mechanism, in which oxygen vacancies are trapped by spin–polarized electrons.[24] These oxygen vacancies with the electrons (acting as F–centers) overlap with $d$–electron spins, which can lead to the FM coupling between the $d$–spins. This mechanism has been applied to explain the magnetism of Co–doped ZnO.[25,26] We postulate that a similar mechanism might be applicable to our sample as well, although we do not yet have a clear evidence for the existence of oxygen vacancies in our system.

On the other hand, we have seen a large AF $\theta_{CW}$ (~ - 320 and -410 K) at high–temperature, but it orders ferromagnetically at 5 K at low–$T$. In fact, this result is similar to that obtained from the Fe–doped BaTiO$_3$,[27] which showed FM signatures at RT with a very large AF correlations. Comparing our results with those obtained from the Fe–doped BaTiO$_3$, it is plausible that the large AF correlations at high temperatures might play a necessary role for the low–$T$ ferromagnetism found in the DMS materials.

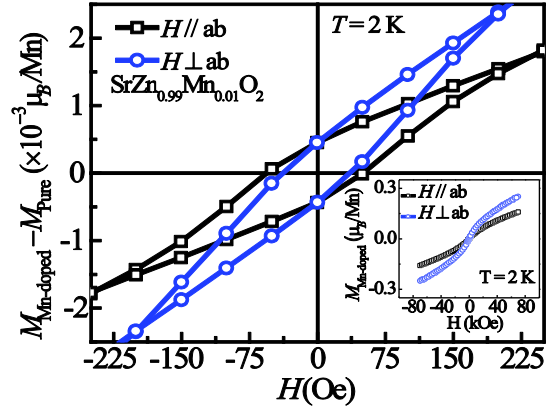

FIG. 5.(color online)$M_{SZM}$–$M_{pure}$ vs $H$ curve of SrZn$_{0.99}$Mn$_{0.01}$O$_2$ at 2K along $H//ab$ (open square) and $H\perp ab$ plane (open circle). The inset shows $M_{SZM}$ vs $H$ curve at 2 K in the field range from –7 T to 7 T for $H//ab$ (open square) and $H\perp ab$ plane (open circle).

If the above scenario is the origin of the weak ferromagnetism in the Mn–doped SrZnO$_2$ sample, then the question arises on why $T_C$ is much lower than that of the Fe–doped BaTiO$_3$. We argue it might be related to the dimensionality of the material. BaTiO$_3$ is a 3–dimensional (3D) material, while SrZnO$_2$ is a quasi–2D material. In general, the 3D interactions are more favorable to form a ferromagnetic or antiferromagnetic state, whereas the 2D systems favors disordered ground states due to the presence of strong quantum fluctuations.[28] Moreover, by doping Fe$^{3+}$ at Ti$^{4+}$ site, it can generate a large number of hole carriers, which seem to be a prerequisite for the RT ferromagnetism. Hence there is a good possibility to have a higher $T_C$ when the SZM system is doped intentionally with carriers either by holes or electrons or by oxygen defects. Note that there is also an example in the family of 3D systems, Mn–doped $SrTiO_3$, in which ferromagnetism is induced at RT upon oxygen defects being introduced.[29]

### 3. $SrZn_{0.99}Cu_{0.01}O_2$ (SZC) sample

Similar to the Mn–doped samples, we have extracted the spin susceptibility of $SrZn_{0.99}Cu_{0.01}O_2$ (SZC) single crystal by subtracting the $\chi_{pure}$ of $SrZnO_2$ from the measured $\chi_{SZC}$ data (for both $H//ab$ and $H\perp ab$). The inversed spin susceptibilities, $(\chi_{SZC}-\chi_{pure})^{-1}$, are plotted as a function of $T$ from 2 to 300 K as shown in Fig. 6. The plot shows the presence of an anisotropy for the data of $H//ab$ and $H\perp ab$, which seems to be associated with the crystallographic anisotropy in this quasi-2D material. No transition is found down to 2 K, unlike the FM-like order observed in the Mn–doped crystals. The data is fitted to the Curie–Weiss law in the $T$–range from 2 to 300 K and confirmed that the SZC is purely paramagnetic down to 2 K. The fitting parameters are listed in Table I. The Curie–constant for $H//ab$ plane give the effective moment of about 1.73 $\mu_B$/Cu as expected for $S = \frac{1}{2}$ moment, but for $H\perp ab$ plane, it is about 1.1 $\mu_B$/Cu, which is slightly smaller than that of $S = \frac{1}{2}$. The inset of Fig. 6 shows the $M$ vs $H$ loop in the field range from –7 T to +7 T at 2 K for $H//ab$ and $H\perp ab$. No hysteresis is found, constituting another evidence for the absence of the FM ordering. The paramagnetic moment of SZC are saturated to 0.95 $\mu_B$/Cu and 0.52 $\mu_B$/Cu for $H//ab$ and $H\perp ab$ plane, respectively. These saturation moment values are consistent with those of effective moment values obtained from the Curie–Weiss fit. The difference in the parameters for $H//ab$ and $H\perp ab$ might be due to the presence of the 2D anisotropy.

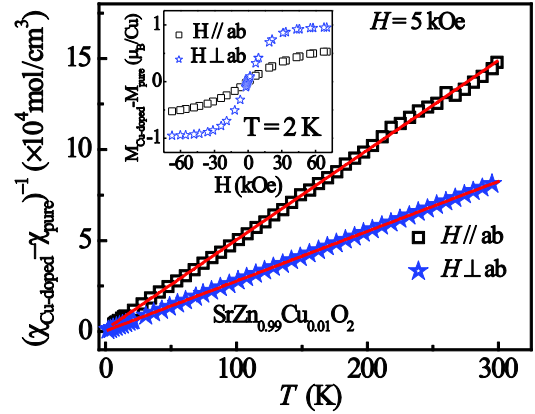

FIG. 6. (color online)The inverse–spin susceptibilities of $SrZn_{0.99}Cu_{0.01}O_2$ $(\chi_{SZC}-\chi_{pure})^{-1}$ vs $T$ for $H//ab$ (open square) and $H\perp ab$ plane (open stars) with Curie–Weiss fits. Inset shows $M$ vs $H$ curve at 2 K.

### IV. Conclusion

In summary, the single crystals of pure, Mn–and Cu–doped $SrZnO_2$ single crystals have been grown by use of an optical floating zone furnace. The 1% Mn-doped sample shows a weak ferromagnetic signature below 5 K, while there exists a large antiferromagnetic coupling with the $\theta_{CW} \approx -320$-410 K at high temperatures. This suggests that the weak ferromagnetism might be associated with the canting of the antiferromagnetically coupled spins. In contrast, the Cu–doped sample does not have any magnetic transition down to 2 K.

### Acknowledgements:

The authors acknowledge the support from National Science Council of Taiwan under project number NSC-101-2119-M-002-007. The work at SNU was supported by the

**Figure caption**

FIG. 1. (color online) The crystal structure of $SrZnO_2$ viewed along the crystallographic $a$–axis. The $ZnO_4$ unit, Sr, and O atoms are represented by green, violet and red colors respectively.

FIG. 2. (color online) X-ray diffraction patterns of the cleaved (00$l$) plane of (a) $SrZnO_2$ (b) 1% Mn–doped and (c) 1% Cu–doped crystals. The color of the crystal changes drastically after even 1% of doping.

FIG. 3. (color online) The spin susceptibilities $\chi$ of $SrZnO_2$ are plotted versus temperature and fitted to the Curie–Weiss law (red line). The inset shows the $M$ vs $H$ curve for the $SrZnO_2$ at 2 K in the $H$–rang from −7 T to +7 T.

FIG. 4. (color online) The inverse of spin susceptibilities, $(\chi_{SZM}-\chi_{pure})^{-1}$, for single crystal $SrZn_{0.99}Mn_{0.01}O_2$ (SZM) along $H//ab$ (open square) and $H\perp ab$ (open circle). Red lines represent the fit to the Curie–Weiss law. The inset shows the data of $\chi_{SZM}$ vs. $T$ in semi–log scale for the $H//ab$ plane of $SrZn_{0.99}Mn_{0.01}O_2$ single crystal at 50 Oe.

FIG. 5. (color online) $M_{SZM}-M_{pure}$ vs $H$ curve of $SrZn_{0.99}Mn_{0.01}O_2$ at 2 K along $H//ab$ (open square) and $H\perp ab$ plane (open circle). The inset shows $M_{SZM}$ vs $H$ curve at 2 K in the field range from −7 T to 7 T for $H//ab$ (open square) and $H\perp ab$ plane (open circle).

FIG. 6. (color online) The inverse–spin susceptibilities of $SrZn_{0.99}Cu_{0.01}O_2$ $(\chi_{SZC}-\chi_{pure})^{-1}$ vs $T$ For $H//ab$ (open square) and $H\perp ab$ plane (open stars) with Curie–Weiss fits. Inset shows $M$ vs $H$ curve at 2 K.